\documentclass[twocolumn,english,prb,showpacs]{revtex4}

\usepackage[T1]{fontenc}
\usepackage[latin9]{inputenc}
\usepackage{float}
\usepackage{amsmath}
\usepackage{graphicx}
\usepackage{amssymb}
\usepackage{esint}
\usepackage{color}
\usepackage{babel}

\begin{document}

\title{Enhanced shot noise in asymmetric interacting two level systems}

\author{Assaf Carmi,$^{1, 2}$ and Yuval Oreg$^1$\\
\textit{$^1$Department of Condensed Matter Physics, Weizmann Institute
of Science, Rehovot, 76100, Israel\\$^2$Department of Particle Physics and Astrophysics, Weizmann Institute
of Science, Rehovot, 76100, Israel}}
\begin{abstract}
We study a model of two interacting levels that are attached to two electronic leads, where one of the levels is attached very weakly to the leads. We use rate equations to calculate the average current and the noise of electrons transmitted through the two levels. We show that the shot noise is enhanced because of the interactions and that the Fano factor depends on the properties of the couplings between the levels and the leads. We study both sequential tunneling and cotunneling processes and show that there is a range of parameters in which the cotunneling processes affect the noise significantly, even though most of the current is carried by sequential tunneling processes.
\end{abstract}


\maketitle

\section{Introduction}
Fluctuations in electrical current, which we simply call \emph{noise} provide additional information about the charge transport that is not accessible from average current measurements (for a review see Ref. [\onlinecite{blanter-2000-336}]). Among the various noise sources we focus on the shot noise. The discreteness of the transferred charge causes fluctuations in the current. These fluctuations, named \emph{shot noise}, depend on the charge of the conducting particles and therefore measurements of shot noise provide information on the discrete nature of the conducting particles and their correlations. The shot noise of a Poisson process of uncorrelated current pulses of charge $e$ is $S_{\rm Shot}=2e\langle I\rangle$, where $\langle I\rangle$ is the time averaged current. The noise is proportional to the average current since in a Poisson distribution the mean equals the variance, therefore the mean number of current pulses equals its variance. The Fano factor, $F=\frac{S_{\rm shot}}{2e\langle I\rangle}$, is a dimensionless parameter that characterizes the granularity of the current. When driving a current through a single spinless electronic level, the correlations that are imposed by the Pauli exclusion principle reduce the Fano factor; the Fano factor varies between half and one depending on the symmetry of the couplings between the level and the external leads \cite{PhysRevB.46.4714,PhysRevB.47.1967,PhysRevB.48.17209}.

Correlations can also be imposed by Coulomb interactions\cite{PhysRevB.63.125315}. In most cases, the effect of Coulomb repulsion on the noise of a mesoscopic system is similar to the effect of the Pauli exclusion principle. Both impose a time delay between consecutive current pulses and therefore we expect negative correlations between them that suppress the shot noise\cite{PhysRevLett.75.3340}. However, several theoretical works on various systems have showed that Coulomb interactions might also lead to a super-Poisson shot noise with a Fano factor $F>1$. Examples of such systems are quantum dots that are coupled to ferromagnetic leads \cite{PhysRevB.60.12246,PhysRevB.62.1186,PhysRevB.76.155408,0953-8984-19-9-096208}, multi-levels quantum dots \cite{PhysRevB.71.045341,PhysRevB.71.161301}, multi-dots structures \cite{Appl.Phys.Lett.87.032105,Appl.Phys.Lett.89.052101,PhysRevLett.99.206602,PhysRevB.77.035302,PhysRevB.77.035409,PhysRevB.78.125308}, and also three terminal quantum dots \cite{PhysRevLett.92.206801,0295-5075-66-3-405,PhysRevB.70.115315}. There are also experimental works \cite{PhysRevLett.91.136801,PhysRevB.73.035424,PhysRevB.74.195305,PhysRevLett.96.246804,PhysRevLett.98.066801} in which a super-Poisson noise was measured in quantum dots, rather than the sub-Poisson noise, which is expected from the single level model of the quantum dots.

A simple mechanism that might explain the enhancement of the Fano factor in such systems is tunneling through two levels that are coupled to the leads, where one of the levels is coupled much stronger than the other level. The two levels are interacting, namely, there is a Coulomb repulsion between electrons that occupy the two levels. In this case, the electrons that cross the system tunnel mainly through the level that is strongly coupled to the leads. However, once in a while an electron can tunnel into the weakly coupled level and then, because of the Coulomb interactions, the tunneling of other electrons through the strongly coupled level is prevented, and the current is blocked. The current resumes only after the electron tunnels out of the weakly coupled level. Therefore the intuitive picture is a current that is blocked occasionally and therefore the noise is enhanced. The idea of two interacting levels as a possible source of super-Poisson noise was discussed in the context of quantum dots that are coupled to ferromagnetic leads\cite{PhysRevB.60.12246,PhysRevB.62.1186} and also in the context of double quantum dots structures \cite{Appl.Phys.Lett.87.032105}.

In this work, we analyze in detail the model of two levels with Coulomb interactions that are attached to two electronic leads. In particular, we study how the shot noise and the Fano factor depend on the left-right asymmetry, namely, the asymmetry between the couplings to the two external leads. The fact that the enhancement of the shot noise depends on the left-right asymmetry of the coupling to the leads emerges from previous works (\emph{e.g.}, the results of Refs. \onlinecite{0953-8984-19-9-096208} and \onlinecite{Appl.Phys.Lett.87.032105}), nevertheless a complete theoretical analysis of this dependence is missing. We find that the Coulomb interactions affect significantly the dependence of the Fano factor on the left-right asymmetry. We study how the temperature affects this dependence and show that finite temperature suppresses the Fano factor at very asymmetric couplings. In addition, we include in our calculations not only the sequential tunneling processes (leading order perturbation in the tunneling coefficients) but also the cotunneling processes (next leading order perturbation) and we find that the latter play an important role in this system. Surprisingly, and perhaps this is the main point of our work, there is a range of parameters where the noise is governed by the rare cotunneling processes even though most of the average current is carried by sequential tunneling processes. The importance of the cotunneling processes follows the fact that they allow the system to change its occupation from one level to the other level in a single quantum process.
The two levels mechanism for noise enhancement can be found in many physical realizations such as single level with spin dependant coupling and double quantum dots or two levels in a quantum dot in a strong magnetic field. Although the quantitative details of each system are different, the qualitative behavior of the noise enhancement is the same.

The paper is organized as follows: in Sec. \ref{sec:The model}, we present the model that we analyze, then we give in Sec. \ref{sec:qualitative description} a simplified qualitative description of the transport through the two levels, which we use to develop an intuitive understanding of the results that we obtain later through the rate equations method. Next, in Sec. \ref{sec:rate equation formalism}, we give the details of the theoretical calculations. In Sec. \ref{sec:results}, we present the results of our calculations and discuss them. Last, we summarize our results in Sec. \ref{sec: conclusions}. In the two appendices, we list all the tunneling rates that are relevant for the theoretical calculations (Appendix \ref{app: list of rates}) and give technical details about the regularization procedure that we use in the calculations of the cotunneling processes' rates (Appendix \ref{app:regularization}).

\section{Model}\label{sec:The model}
\begin{figure}
\includegraphics[bb=115bp 200bp 530bp 490bp,clip,scale=0.5]{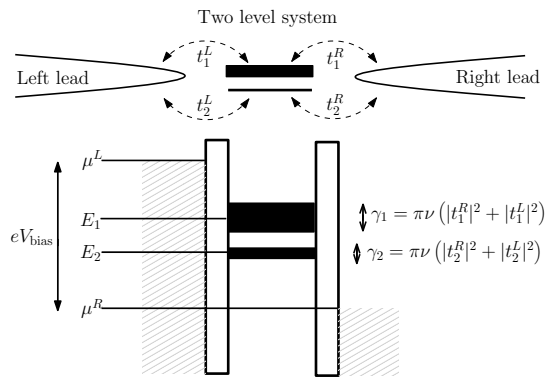}
\caption{A schematic picture of the model. The energies of the two levels are $E_1$ and $E_2$. The bias voltage is the difference between the two chemical potentials: $eV_{{\rm bias}}=\mu^{L}-\mu^R$. Here the width of level 2 is much smaller than the width of level 1 which is assumed to be much smaller than the bias voltages: $\gamma_2\ll\gamma_1\ll|\mu^{R,L}-E_{1,2}|$. We assume that the tunneling coefficients are independent of the energy and we also assume for simplicity that they all have the same sign, and that the two ratios between the left tunneling coefficients and the right tunneling coefficients are identical, $t_1^L/t_1^R=t_2^L/t_2^R$.}
\label{fig:system}
\end{figure}

We consider a model of two interacting levels that are attached to two leads (see Fig. \ref{fig:system}). For simplicity, we discuss a spinless problem\cite{footnote_on_spin}. One of the levels (level 2 in Fig. \ref{fig:system}) is attached very weakly (compared to the other level) to the leads. A simultaneous occupation of the two levels is possible, however this situation is not likely to happen as it requires an additional charging energy, $U$, because of the Coulomb interaction. This model describes, for example, an interacting two level quantum dot in a strong magnetic field. The Hamiltonian that describes the system is
\begin{equation}
H=H_{\rm leads}+H_{\rm 2 levels}+H_{\rm t},\label{eq:model_Hamiltonian}
\end{equation}
where
\begin{eqnarray*}
&H_{\rm leads}&=\sum_k\epsilon^L_{k}L^{\dagger}_{k}L_{k}+\sum_k\epsilon^R_{k}R^{\dagger}_{k}R_{k}, \\
&H_{\rm 2 levels}&=E_{1}d^{\dagger}_{1}d_{1}+E_{2}d^{\dagger}_{2}d_{2}+Ud^{\dagger}_{1}d_{1}d^{\dagger}_{2}d_{2}, \\
&H_{\rm t}&=\sum_{i,k}\left(t^{L}_{i}L^{\dagger}_{k}d_{i}+t^{R}_{i}R^{\dagger}_{k}d_{i}\right)+H.c.\mbox{\ \ .}
\end{eqnarray*}
Here, $L_k$ ($R_k$) are left (right) lead annihilation operators, $d_i$ is the $i$th level annihilation operator, and we have assumed that the tunneling coefficients ($t^{L,R}_{i}$) are independent of the energy. We also assume that all the tunneling coefficients have the same sign.

We calculate the average current and the zero-frequency noise using rate equations method in which we assume that the two levels are weakly coupled to the leads. By weakly coupled we mean that the widths of the levels are much smaller than the temperature, or alternatively, at low temperatures it means that the bias voltage (see Fig. \ref{fig:system}) is much larger than the levels' widths
\begin{equation}
\gamma_i=\pi\nu\left(\left|t^L_i\right|^2+\left|t^R_i\right|^2\right)\ll\left|\mu^{R,L}-E_i\right|,\label{eq: definition of width}
\end{equation}
where $\mu^{R,L}$ are the electro-chemical potentials of the leads and we have assumed for simplicity the same density of states, $\nu$, in the two leads.
\section{Qualitative simplified model}\label{sec:qualitative description}
In this section, we study a simplified intuitive model that captures, at least qualitatively, most of the results that we later achieve through a more rigorous analysis. The propose of this section is to establish a simple physical picture that we can use to interpret the results that we get through the rate equations formalism. Consider the two level system that was discussed in the previous section and depicted in Fig. \ref{fig:system} and assume strong interactions, \emph{i.e.}, a large $U$. At this point, we also assume, for simplicity, zero temperature. As a function of $E_1$, there are two regions where we expect to have current: at $\mu^R<E_1<\mu^L$ where the two levels are between the chemical potentials (assuming that $|E_1-E_2|\ll V_{\rm bias}$), and at $\mu^R<E_1+U<\mu^L$ where one of the levels is occupied and effectively, because of the Coulomb interaction, the other level is shifted up and placed between the chemical potentials. We choose to focus in this section on the later: $\mu^R<E_1+U<\mu^L$.

Since level 2 is coupled weaker than level 1 to the leads, most of the time the current flows through level 1 (\emph{i.e.} level 2 is occupied and electrons enter and leave level 1). However, after a while, the electron in level 2 can tunnel out to the right lead and by that, because of the Coulomb interaction, it reduces the effective energy of level 1 (from $E_1+U$ to $E_1$) making the tunneling out of level 1 impossible. The current is therefore blocked. The current resumes only after a new electron from the left lead tunnels into level 2. Hence, the picture is a current (through level 1) that is stopped occasionally (by tunneling out of level 2). This situation is schematically drawn in Fig. \ref{fig:qualitative}.

\subsection{Simplified model: shot noise on top of a telegraph noise}
\begin{figure}
\includegraphics[bb=45bp 180bp 390bp 390bp,clip,scale=0.7]{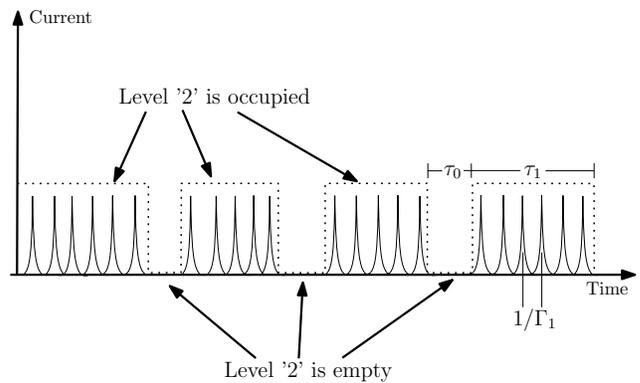}
\caption{A qualitative description of the transport through the two level system in the region $\mu^R<E_1+U<\mu^L$. Electrons are tunneling through level 1 until the electron in level 2 tunnels out and then the current is blocked. The tunneling events through level 1 resume when another electron tunnels into level 2. The current behaves as a multiplication of two signals: A sequence of current pulses through level 1 and a telegraphic signal, the occupation of level 2. $\Gamma_1$ is the average rate of tunneling events through level 1, $\tau_{\rm 0}$ is the average time duration until level 2 is filled, and $\tau_{\rm 1}$ is the average time duration until level 2 gets empty.}
\label{fig:qualitative}
\end{figure}

The current, as drawn in Fig. \ref{fig:qualitative}, fluctuates between two modes: A zero-current mode where the current is dramatically suppressed, and a nonzero mode where the current is carried by pulses of charge (electrons) that tunnel through level 1. We therefore suggest the following simplified model: the current $I(t)$ is a multiplication of two signals
\begin{equation}
I(t)=I_{\rm shot}(t)\times C_{\rm telegraph}(t),\label{eq:simplified_current}
\end{equation}
where $I_{\rm shot}(t)$ is the current through level 1 and $C_{\rm telegraph}(t)$ is a random telegraph signal that fluctuates between two values, zero and one, according to the occupation of level 2. In this simplified model, the current flows only through level 1 and only when level 2 is occupied.
We neglect the effects that the Pauli principle imposes on the current $I_{\rm shot}(t)$ and treat it as a sequence of current pulses of charge $e$ with a Poissonian statistics characterized by a rate $\Gamma_1$. The changes in the value of $C_{\rm telegraph}$ are Poissonian events with the rates $1/\tau_{1}$ and $1/\tau_{0}$ for the events of changing the value of $C_{\rm telegraph}$ from one to zero and zero to one, respectively. The fluctuations in the current $I_{\rm shot}(t)$ are known as shot noise, and the fluctuations in the signal $C_{\rm telegraph}(t)$ are known as telegraph noise \cite{machlup:341}.

Assuming that $I_{\rm shot}$ and $C_{\rm telegraph}$ are uncorrelated, the average current is
\begin{equation}
\langle I\rangle=\langle I_{\rm shot}\rangle\langle C_{\rm telegraph}\rangle=e\Gamma_1\frac{\tau_{1}}{\tau_{1}+\tau_{0}}.
\end{equation}
The noise is related to the auto-correlation function of the total current via Wiener-Khinchin theorem
\begin{eqnarray}
S(\omega)&=&2\int_{-\infty}^\infty d\tau e^{i\omega\tau}(\langle I(t+\tau)I(t)\rangle-\langle I\rangle^2)\\
&=&\langle I_{\rm shot}\rangle^2S_{\rm telegraph}(\omega)+\langle C_{\rm telegraph}\rangle^2S_{\rm shot}(\omega)\nonumber\\
&+&\frac{1}{\pi}S_{\rm shot}(\omega)\ast S_{\rm telegraph}(\omega)\nonumber.
\end{eqnarray}
Using the known results for the shot noise and the telegraph noise \cite{machlup:341} we get
\begin{equation}
S(\omega)=2e^2\Gamma_1\frac{\tau_{1}}{\tau_{1}+\tau_{0}}+\frac{4e^2\Gamma_1^2}{(\tau_{1}+\tau_{0})((1/\tau_{1}+1/\tau_{0})^2+\omega^2)}.\label{eq:simplified_model_noise}
\end{equation}
\subsection{Fano factor and asymmetry dependence}
Before we study the Fano factor and its left-right asymmetry dependence, it seems necessary to inquire about the linear dependence of the noise on the current as the second term in (\ref{eq:simplified_model_noise}) is quadratic in $\Gamma_1$. Since the rates $1/\tau_{1}$ and $1/\tau_{0}$ are the rates of tunneling out of and into level 2, they depend, up to a symmetry factor, linearly on $\gamma_2$ that was defined in Eq. (\ref{eq: definition of width}). Similarly, $\Gamma_1$ is linear in $\gamma_1$. Therefore, under the reasonable assumption that $\gamma_1/\gamma_2$ is independent of the left-right asymmetry (that is, the couplings of the levels to the leads cannot be changed independently), $1/\tau_{1}$ and $1/\tau_{0}$ are linear in $\Gamma_1$. Hence, the zero-frequency noise, $S(0)$, depends linearly on $\Gamma_1$ and therefore depends linearly on the current.

The probability of finding level 2 occupied, which is the probability of finding $C_{\rm telegraph}=1$, is $P_{\rm f}=\frac{\tau_1}{\tau_0+\tau_1}$. Similarly, the probability of finding level 2 empty, which at zero temperature is the probability of finding \emph{only} level 1 occupied is $P_{\rm e}=\frac{\tau_0}{\tau_0+\tau_1}$. The Fano factor is
\begin{equation}
F=\frac{S(0)}{2e\langle I\rangle}=1+\frac{2\tau_0^2\tau_1\Gamma_1}{(\tau_0+\tau_1)^2}=1+\Gamma_1\tau_1 2P_{\rm e}^2.\label{eq:Fano_simplified_model}
\end{equation}
Notice that $\Gamma_1\tau_1$ is the average number of tunneling events during a $C_{\rm telegraph}=1$ stage, which is the average number of tunneling events through level 1 before a tunneling event out of level 2 takes place. At zero temperature it can be estimated as $|t_1^R|^2/|t_2^R|^2$, or, assuming the same left-right asymmetry for the two levels, $\Gamma_1\tau_1=\gamma_1/\gamma_2$. The Fano factor (\ref{eq:Fano_simplified_model}) becomes
\begin{equation}
F=1+2\frac{\gamma_1}{\gamma_2}P_{\rm e}^2.\label{eq:Fano_simplified_zero_T}
\end{equation}

The probability $P_{\rm e}$ depends on the asymmetry between the coupling to the left and the right leads, \emph{i.e.} on the ratio $|t_i^L|^2/|t_i^R|^2$. To see this consider the simple case of zero temperature and large U. For $\mu^L>\mu^R$ (see Fig. \ref{fig:system}), the rate of tunneling out of level 2 is, according to Fermi's golden rule, $\Gamma_{\rm f \rightarrow e}=\frac{2\pi}{\hbar}|t_2^R|^2$. Similarly, the rate of tunneling into level 2 is $\Gamma_{\rm e \rightarrow f}=\frac{2\pi}{\hbar}|t_2^L|^2$. If we neglect cotunneling effects there are no direct tunneling processes from level 1 to level 2, so the steady state probabilities satisfy
\begin{equation}
P_{\rm e}\Gamma_{\rm e \rightarrow f}=P_{\rm f}\Gamma_{\rm f \rightarrow e}.\label{eq: steady state solution}
\end{equation}
$P_{\rm f}$ is the probability of finding both level 1 and level 2 occupied. In the limit $|E_1-E_2|\ll V_{\rm bias}$, the probability of finding only level 1 occupied, $P_{\rm e}$, equals the probability of finding only level 2 occupied. In the limit of large $U$, the probability of finding both level 1 and level 2 empty (for $\mu^R<E_1+U<\mu^L$) is zero. Thus, $P_{\rm f}+2P_{\rm e}=1$. The steady state solution (\ref{eq: steady state solution}) becomes
\begin{equation}
P_{\rm e}=\frac{\Gamma_{\rm f \rightarrow e}}{\Gamma_{\rm e \rightarrow f}+2\Gamma_{\rm f \rightarrow e}}=\frac{1}{|t_i^L|^2/|t_i^R|^2+2}.\label{eq: solution of Pe}
\end{equation}
By increasing $|t_i^L|^2/|t_i^R|^2$, we decrease $P_{\rm e}$. In the limit $|t_i^L|^2/|t_i^R|^2\rightarrow\infty$, $P_{\rm e}\rightarrow0$ so the Fano factor (\ref{eq:Fano_simplified_zero_T}) $F\rightarrow1$. By decreasing $|t_i^L|^2/|t_i^R|^2$ we increase $P_{\rm e}$ and in the limit $|t_i^L|^2/|t_i^R|^2\rightarrow0$ it gets its maximal value $P_{\rm e}=1/2$ and the Fano factor is maximal, $F=1+\frac{\gamma_1}{2\gamma_2}\approx\frac{\gamma_1}{2\gamma_2}$.

\subsubsection{Finite temperature}
At zero temperature, the only possible tunneling event that follows a tunneling event from the system to the right lead, is from the left lead into the empty level.
Therefore, changing the value of $C_{\rm telegraph}$ from one to zero in the simplified model, corresponds to a tunneling event from level 2 to the right. In addition, every tunneling event from level 1 to the right, corresponds to a current-pulse in the simplified model. Hence, we can estimate the number of tunneling events during a $C_{\rm telegraph}=1$ stage as $\Gamma_1\tau_1\approx|t_1^R|^2/|t_2^R|^2$. At finite temperature, there is also a finite probability that a tunneling event from the system to the right will be followed by a tunneling event from the right lead back to the empty level. In the limit $|t_i^L|^2/|t_i^R|^2\rightarrow0$ the probability of tunneling from right to left may become important. If the couplings to the right are much stronger than the couplings to the left, electrons may tunnel many times back and forth between the right lead and the system before a tunneling event from the left lead to the system takes place. The time scales $\tau_1$ and $\tau_0$ in this case, become smaller than $1/\Gamma_1$, and the value of $\Gamma_1\tau_1$ goes to zero. Therefore, we expect that at finite temperature the Fano factor will have the value one in the limit $|t_i^L|^2/|t_i^R|^2\rightarrow0$. To conclude, as we decrease $|t_i^L|^2/|t_i^R|^2$ the Fano factor (\ref{eq:Fano_simplified_model}) gets larger, but at finite temperature, at some point, if we decrease $|t_i^L|^2/|t_i^R|^2$ even further, the Fano factor will start to decrease toward the value one at $|t_i^L|^2/|t_i^R|^2\rightarrow0$.

\subsubsection{Cotunneling effects}
Similar to finite temperature, cotunneling processes may suppress the Fano factor at $|t_i^L|^2/|t_i^R|^2\rightarrow0$. Consider the limit $|t_i^L|^2\ll|t_i^R|^2$. Without cotunneling, if level 2 is empty, the electron in level 1 need to wait a long time before it can tunnel out to the right lead since such a tunneling event must follows a tunneling event into level 2 which, at zero temperature, is possible only from the left lead. Cotunneling processes, however, allow the two processes at once; occupying level 2 and evacuating level 1 in a single quantum process. In particular, the electron in level 1 can virtually tunnel out to the right lead while another electron is virtually tunnel from the right lead into level 2 (we use the term virtually to emphasize the fact that the intermediate state does not conserve energy). In the limit $|t_i^L|^2/|t_i^R|^2\rightarrow0$ the total rate of such processes may become larger than the rate of sequential tunneling from the left lead into level 1. Thus the occupation of level 2 and therefore the telegraphic signal, $C_{\rm telegraph}(t)$, fluctuates much faster than the pulses' rate $1/\Gamma_1$, and the value of $\Gamma_1\tau_1$ goes to zero. Therefore we expect a suppression of the Fano factor due to the cotunneling processes toward the value one in the limit $|t_i^L|^2/|t_i^R|^2\rightarrow0$.
\subsection{Weak interactions}
Up to this point, we have assumed strong interactions, namely, a very large $U$. If $U$ is not large compared to the bias voltage, $U<eV_{\rm bias}$, changing the occupation of level 2 doesn't block completely the current through level 1 since electrons can tunnel through level 1 in both cases; while level 2 is empty or occupied. Yet, we can still use the intuitive picture of a sequence of current pulses through level 1 and a random telegraph signal describing the occupation of level 2. The rate of the pulses depends on the occupation of level 2 and we consider two different rates: $\Gamma_1$ describes the tunneling rate through level 1 while level 2 is full and $\tilde{\Gamma}_1$ describes the tunneling rate through level 1 while level 2 is empty (previously $\tilde{\Gamma}_1$ was zero). Similar to Eq. (\ref{eq:simplified_current}), we consider the current
\begin{equation}
I(t)=I_{\rm shot}(t)\times C_{\rm telegraph}(t)+\tilde{I}_{\rm shot}(t)\times (1-C_{\rm telegraph}(t)),
\end{equation}
where $I_{\rm shot}$ ($\tilde{I}_{\rm shot}$) is a sequence of current pulses with a characteristic rate $\Gamma_1$ ($\tilde{\Gamma}_1$). The Fano factor is
\begin{equation}
F=1+\frac{2(\Gamma_1-\tilde{\Gamma}_1)^2\tau_0^2\tau_1^2}{(\tau_0+\tau_1)^2(\tau_1\Gamma_1+\tau_0\tilde{\Gamma}_1)},\label{eq:Fano_simplified_smallU}
\end{equation}
where $\tau_{1}$ is the average time duration in which level 2 is occupied and electrons tunnel through level 1 with an average rate $\Gamma_1$, and $\tau_{\rm 0}$ is the average time duration in which level 2 is empty and electrons tunnel through level 1 with an average rate $\tilde{\Gamma}_1$.

At low temperatures, $K_BT\ll eV_{\rm bias}$, if $\mu^R<E_1<\mu^L$ and $\mu^R<E_1+U<\mu^L$, the tunneling rate through level 1 barely depends on the occupation of level 2, \emph{i.e.} $\Gamma_1\approx\tilde{\Gamma}_1$. We can approximate (\ref{eq:Fano_simplified_smallU}) by
\begin{equation}
F\approx 1+\frac{2(\Delta\Gamma_1)^2}{\Gamma_1\Gamma_2}P_{\rm e}^2(1-P_{\rm e})^2,\label{eq:Fano_simplified_smallU_approx}
\end{equation}
where $\Delta\Gamma_1\equiv\tilde{\Gamma}_1-\Gamma_1$ and $\Gamma_2=(\tau_0+\tau_1)^{-1}$ (at low temperatures this is the tunneling rate through level 2). Notice that $P_{\rm e}=\frac{\tau_0}{\tau_0+\tau_1}$, which is the probability of finding level 2 empty, is different from the probability of finding only level 1 occupied, since now the two levels can be empty simultaneously.

The results of the simplified model which we presented in this section, can be used to analyze transport through other systems as well. Systems with discrete charge carriers and in addition a telegraphic behavior can be analyzed in the same way. Example of such a system is a quantum point contact capacitively coupled to a system that alternates between two charge states \cite{PhysRevLett.100.176805}.
\section{Rate equations formalism}\label{sec:rate equation formalism}
The two level system is weakly coupled to the leads, therefore we may use the rate equations formalism \cite{PhysRevB.49.10381} to calculate the average current and the current noise. In the weak coupling regime, we describe the whole system using the four eigenstates of the two level system: $|n_1,n_2\rangle$ ($n_i$ labels the eigenvalues of $d^{\dagger}_id_i$) and treat the tunneling Hamiltonian, $H_{{\rm t}}$, as a perturbation. Transition rates between these states are related to rates of electron tunneling events. In the following two subsections we first review the method of current and noise calculations using rate equations and we then discuss the tunneling processes that are relevant to our model.
\subsection{Rate equations}
We shortly review the technique for calculating current and shot noise that was developed in Ref. [\onlinecite{PhysRevB.49.10381}] and the generalization of it to include cotunneling processes that was developed in Ref. [\onlinecite{PhysRevB.74.205438}]. Consider the two level system that we discuss. The system can be in one of four states, $|n_1,n_2\rangle$, and its dynamics is driven by tunneling processes. Generally, there are a few tunneling processes that change the system's state from $|i\rangle$ to $|f\rangle$ which we denote their rates by $\omega_{i,f}^{\alpha}$ where $\alpha$ labels the different possible tunneling processes. The rate $\omega_{i,f}=\sum_{\alpha}\omega_{i,f}^{\alpha}$ is the total transition rate from the state $|i\rangle$ to the state $|f\rangle$.
The system's dynamics is described by the rate equations
\begin{equation}
\frac{\partial}{\partial t}P(f,t/i)=\sum_{k\in S}\left(P(k,t/i)\omega_{k,f}-P(f,t/k)\omega_{f,k}\right),
\label{eqn:master}
\end{equation}
where $P(f,t/i)$ is the probability of the system to be in the state $|f\rangle$ at time $t$ if it was in the state $|i\rangle$ at $t=0$, so the initial condition is $P(f,t=0/i)=\delta_{i,f}$.
Eq. (\ref{eqn:master}) neglects coherence superpositions of different states (the terms $P(f,t/i)$ are the diagonal matrix elements of the system's density matrix). Neglecting the coherent superpositions of the states $|1,0\rangle$ and $|0,1\rangle$ is justified when the coherence time is much shorter than the delay time between consecutive tunneling events. Alternatively, if there is a quantum number that distinguishes the two states (\emph{e.g.}, spin), if at some point in time the density matrix that describes the system is diagonal, then coherent superpositions are zero at all later times.

It appears to be useful to write the rate equations in a matrix form. We define the vector (there are four such vectors)
\begin{equation}
\textbf{P}_{i}(t)\equiv(P(00,t/i),P(10,t/i),P(01,t/i),P(11,t/i)),
\end{equation}
and the matrix
\begin{footnotesize}
\begin{equation}
M=\left(
                                              \begin{array}{cccc}
                                                -\displaystyle\sum_{k\neq00}\omega_{00,k} & \omega_{10,00} & \omega_{01,00} & \omega_{11,00} \\
                                                \omega_{00,10} & -\displaystyle\sum_{k\neq10}\omega_{10,k} & \omega_{01,10} & \omega_{11,10} \\
                                                \omega_{00,01} & \omega_{10,01} & -\displaystyle\sum_{k\neq01}\omega_{01,k} & \omega_{11,01} \\
                                                \omega_{00,11} & \omega_{10,11} & \omega_{01,11} & -\displaystyle\sum_{k\neq11}\omega_{01,k} \\
                                              \end{array}
                                            \right).
\label{eqn:M}
\end{equation}
\end{footnotesize}
The rate equations become $\frac{\partial}{\partial t}\textbf{P}_{i}(t)=M\textbf{P}_{i}(t)$, with the initial condition $\textbf{P}_{i}(t=0)=(0,\ldots,0,1,0,\ldots,0)\equiv\widehat{e}_{i}$.
The solution of Eq. (\ref{eqn:master}) is readily found to be $\textbf{P}_{i}(t)=e^{Mt}\widehat{e}_{i}$.
Let $P^{st}$ be the stationary solution, namely, $MP^{st}=0$ and (since it is a probabilities vector) $\displaystyle\sum_{n}P^{st}_{n}=1$.

{\it current}. We define the quantity $s^{a\alpha}_{i,f}$, the total number of electrons that tunnel through the junction $a=L,R$ to the right during the process $\alpha$ that changes the system's state from $|i\rangle$ to $|f\rangle$. A negative sign corresponds to electrons that are moving to the left. In our model, $s^{a\alpha}_{i,f}$ can get the values $\pm2,\pm1,0$.
The stationary current through the junction $a$ can be written as
\begin{equation}
\langle I_{a}\rangle=e\displaystyle\sum_{i,f\in S}\displaystyle\sum_{\alpha}s^{a\alpha}_{i,f}P^{st}_{i}\omega_{i,f}^{\alpha}.
\label{eq0}
\end{equation}
{\it Zero frequency noise}. The noise is related to the auto-correlation function via Wiener-Khinchin theorem
\begin{equation}
S_{ab}(\omega)=2\int_{-\infty}^\infty d\tau e^{i\omega\tau}(\langle I_{a}(t+\tau)I_{b}(t)\rangle-\langle I\rangle^2),
\end{equation}
where we are interested in the zero frequency limit.
We write here a compact expression for the noise, details of the derivation can be found in Ref. [\onlinecite{PhysRevB.74.205438}]. The zero frequency noise can be written as
\begin{equation}
S_{ab}(\omega\rightarrow0)=2e^2\{tr\textbf{U}_{ab}-\textbf{W}_b\cdot M^{-1}\bar{\textbf{Y}}_a-\textbf{W}_a\cdot M^{-1}\bar{\textbf{Y}}_b\},
\label{eqn:zero-frequency-noise}
\end{equation}
with the following vectors:
\begin{eqnarray}
&(\textbf{U}_{ab})_{i}&\equiv\sum_{f\in S}\sum_{\alpha}s^{a\alpha}_{i,f}s^{b\alpha}_{i,f}P^{st}_{i}\omega^{\alpha}_{i,f}, \\&
(\bar{\textbf{Y}}_{a})_j&\equiv\sum_{i\in S}\sum_{\alpha}s^{a\alpha}_{i,j}P^{st}_{i}\omega^\alpha_{i,j}-\frac{\langle I_a\rangle}{e}P^{st}_j, \\&
(\textbf{W}_{b})_{k}&\equiv\sum_{f\in S}\sum_{\alpha}s^{b\alpha}_{k,f}\omega^{\alpha}_{k,f}.
\end{eqnarray}
We have used the trace of a vector to denote the sum of its elements. Although the matrix $M$ is not invertible, there is only one traceless vector $\textbf{V}_a$ (\emph{i.e.} the sum of all its elements is zero) that satisfies  $M\textbf{V}_a=\bar{\textbf{Y}}_{a}$ \cite{PhysRevB.74.205438} and we use this vector as $ M^{-1}\bar{\textbf{Y}}_{a}$.
\subsection{Tunneling rates}
We describe the system using four states, $|n_1,n_2\rangle$. The system dynamics is driven by transitions between states that are caused by tunneling processes. For example: Transition from the state $|0,0\rangle$ to the state $|1,0\rangle$ happens when an electron is tunneling from the left lead or the right lead into level 1 while level 2 is empty. The term $H_{{\rm t}}$ in the Hamiltonian (\ref{eq:model_Hamiltonian}) allows tunneling processes and the rates of the transitions are derived perturbatively in $H_{{\rm t}}$.

\begin{figure}
\includegraphics[bb=33bp 155bp 535bp 545bp,clip,scale=0.5]{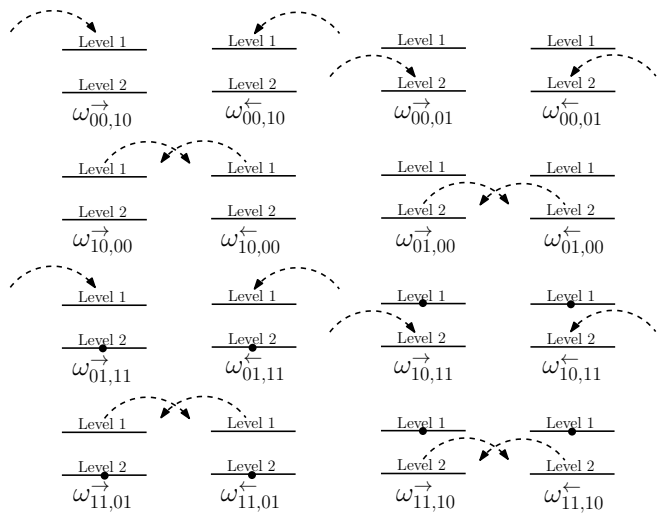}
\caption{A schematic picture of the lowest order in $H_{{\rm t}}$ transitions (the sequential tunneling processes) and their notations.}
\label{fig: sequential processes}
\end{figure}

{\it Sequential tunneling rates}. To the lowest order in $H_{{\rm t}}$ the transition rates can be calculated using Fermi's golden rule. We use the notation $\omega_{i,j}^{\rightarrow}$ for the rate of a tunneling process that changes the system's state from '$i$' to '$j$' by tunneling an electron from the left to the right direction (and similarly $\omega_{i,j}^{\leftarrow}$ for electron that moves from right to left). For example,
\begin{equation}
\omega_{00,10}^{\rightarrow}=\Gamma^L_1 F_{FD}(E_1-\mu^L)
\end{equation}
is the rate of tunneling from the left lead into level 1 while level 2 is empty, where we have defined
\begin{equation}
\Gamma^L_i\equiv\nu\frac{2\pi}{\hbar}|t_i^L|^2,\mbox{\ \ } \Gamma^R_i\equiv\nu\frac{2\pi}{\hbar}|t_i^R|^2,\label{eq: definition_of_Gamma}
\end{equation}
and the Fermi's function $F_{FD}(x)=(1+e^{\beta x})^{-1}$ gives the probability for the availability of an electron for the tunneling process. In Fig. \ref{fig: sequential processes} we depict the lowest order tunneling processes, to which we refer as \emph{sequential tunneling} processes. The rates of all the sequential tunneling processes are listed in Appendix \ref{app: list of rates}.

{\it Cotunneling rates}. The next leading order perturbation in $H_t$ generates cotunneling processes with intermediate virtual states. Usually, when the tunneling coefficients are small ($\gamma_{1,2}$ of Eq. (\ref{eq: definition of width}) are much smaller than the temperature or bias voltage) the higher order perturbation theory is not crucial as it barely improves the approximation. However, this is not always true. Consider, for example, the limit $t_i^R\ll t_i^L$; in this case, a cotunneling process, which takes an electron out of level 2 into the left lead and takes another electron from the left lead into level 1, can become more likely to happen than a sequential tunneling from level 2 to the right lead. In other words, a second leading order perturbation in $t_i^L$ can be more important that a first leading order perturbation in $t_i^R$. We discuss later the importance of the cotunneling processes in the two level model and at the moment we emphasize that this is more than a small improvement of the approximation. We discuss two types of cotunneling processes: elastic-cotunneling, namely, processes that contribute current but don't change the state of the two level system, and inelastic-cotunneling, namely, processes that change the state of the two level system (and might not contribute to the current through it).

\begin{figure}
\includegraphics[bb=5bp 378bp 570bp 710bp,clip,scale=0.45]{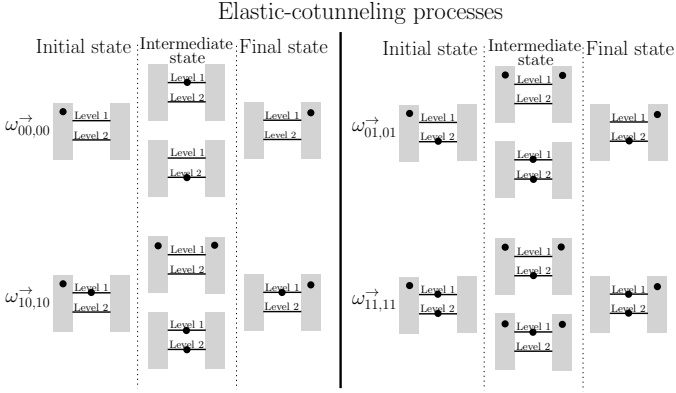}
\caption{A schematic picture of the elastic cotunneling processes and their notations. Each process has two possible virtual intermediate states. The rate is the sum of the amplitudes of the two possible paths (rather than the sum of their probabilities). The processes $\omega_{i,i}^{\leftarrow}$ are the same as $\omega_{i,i}^{\rightarrow}$ after exchanging the final states with the initial states.}
\label{fig: elastic_cotunneling_processes}
\end{figure}

{\it Elastic-cotunneling rates}. Figure \ref{fig: elastic_cotunneling_processes} depicts schematically the elastic-cotunneling processes that we take into account in the transport calculations of the two level system. Each process has two possible intermediate states. For example, electron can tunnel through an empty system via level 1 or 2, thus, cotunneling processes of the form $|0,0\rangle\longrightarrow|0,0\rangle$ have two possible intermediate states: $|1,0\rangle$ and $|0,1\rangle$. The two possible paths interfere and we need to sum the amplitudes of the two possibilities rather than their probabilities. If there is a quantum number that distinguish the two levels (\emph{e.g.}, if the system is a single spinful level with spin-dependent couplings to the leads) the two path do not interfere, and we simply sum their rates. We use the notation $\omega_{i,i}^{\rightarrow}$ ($\omega_{i,i}^{\leftarrow}$) for elastic-cotunneling processes in which an electron is tunneling to the right (left) direction (see Fig. \ref{fig: elastic_cotunneling_processes}). Elastic-cotunneling processes in which an electron tunnels back and forth between one of the levels and one of the leads don't change the state of the system and don't contribute any current, and therefore don't appear directly in the transport calculations. The total elastic-cotunneling rates are the sum of the rates of all the possible processes, namely, integration over all incoming electron's energies. For example,
\begin{eqnarray}
\omega_{00,00}^{\rightarrow}&=&\frac{2\pi\nu^2}{\hbar}\int d\epsilon F_{FD}\left(\epsilon-\mu^L\right)\left(1-F_{FD}\left(\epsilon-\mu^R\right)\right)\nonumber\\&\times&\left|\frac{t^L_1t^R_1}{\epsilon-E_1}+\frac{t^L_2t^R_2}{\epsilon-E_2}\right|^2.\label{eq:elastic_example}
\end{eqnarray}
The rates of all the elastic-cotunneling processes are listed in Appendix \ref{app: list of rates}.
Equation (\ref{eq:elastic_example}) is a formal expression and the actual rate, which we use in the rate equations, cannot be directly calculated from it. The reason is the divergence of the integral due to the finite widths of the two levels (which we treat as two delta functions in energy). This divergence was already discussed before \cite{Averin1994979,PhysRevB.65.115332}, and a regularization scheme for the calculation of the cotunneling rates was developed. We summarize the regularization scheme in Appendix \ref{app:regularization}.

It is worth noting that one can avoid the necessity of regularization by using the diagrammatic technique that was developed in Refs. [\onlinecite{PhysRevLett.76.1715,PhysRevLett.78.4482,PhysRevB.54.16820}]. We find that additional correction due to levels shifts and broadening captured by this approach\cite{PhysRevB.58.7882} are irrelevant in the large bias limit ($\gamma_i\ll eV_{\rm bias}$) that we consider. Calculation procedure of the average current and current noise using this technique was developed in Ref. [\onlinecite{PhysRevB.68.115105}] and gives the same results in the $\gamma_i\ll eV_{\rm bias}$ limit.

\begin{figure}
\includegraphics[bb=0bp 0bp 593bp 842bp,clip,scale=0.45]{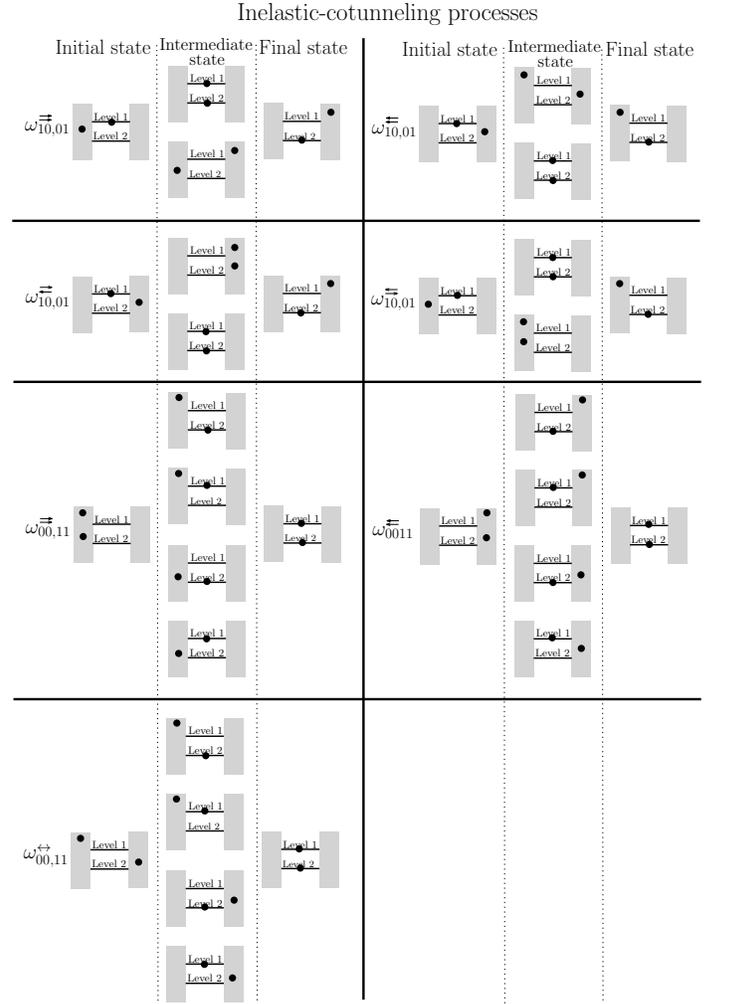}
\caption{A schematic picture of the inelastic cotunneling processes and their notations. The processes $\omega_{01,10}$ are the same as $\omega_{10,01}$ after exchanging the two levels. The processes $\omega_{11,00}$ are the same as $\omega_{00,11}$ after exchanging the final states with the initial states.}
\label{fig: inelastic_cotunneling_processes}
\end{figure}

{\it Inelastic-cotunneling rates}. In Fig. \ref{fig: inelastic_cotunneling_processes}, we depict the cotunneling processes that change the state of the two level system, \emph{i.e.} the inelastic-cotunneling processes. To this order in $H_t$, the inelastic-cotunneling processes change the system's state between the following states: $|1,0\rangle\longleftrightarrow|0,1\rangle$, $|0,0\rangle\longleftrightarrow|1,1\rangle$. The latter are somewhat more complex than the other cotunneling processes as they have four possible intermediate states (see Fig. \ref{fig: inelastic_cotunneling_processes}). We use the notation $\omega_{i,j}^{\rightleftarrows}$ for the rate of processes in which the system changes its state from $|i\rangle$ to $|j\rangle$ in the following way: the electron that enters or leaves level 1 tunnels to the right direction, while the electron that enters or leaves level 2 tunnels to the left, and similarly we define $\omega_{i,j}^{\leftrightarrows},\omega_{i,j}^{\rightrightarrows},\omega_{i,j}^{\leftleftarrows}$. We use the notation $\omega_{00,11}^{\leftrightarrow}$, $\omega_{11,00}^{\leftrightarrow}$ for processes in which the two electrons enter or leave the two levels by tunneling one to the right and the other to the left. The rates of all the inelastic-cotunneling processes are listed in Appendix \ref{app: list of rates}.

\emph{Beyond the cotunneling approximation}. The rate equations based calculation is valid as long as the tunneling coefficients are small enough as we insert the tunneling processes only up to second order in the perturbation $H_t$. Practically, it means that either $\gamma_i/eV_{\rm bias}$ or $\gamma_i/K_BT$ need to be small numbers. Yet, the next leading order in $H_t$ generates logarithmic contributions that diverge at low temperatures and bias voltages\cite{Hewson}. Hence, for bias voltages smaller than a characteristic energy scale, the Kondo temperature, the perturbative approach fails. The Kondo temperature in our case is $T_K\sim\sqrt{\gamma_1U}e^{-U/2\gamma_1}$ and in all cases in this work we consider much larger bias voltages. We also want to note that small corrections due to the renormalization of the energy levels and broadening play very minor role in the large bias case that we consider. They at most slightly modify the quantitative results with no important effect on the qualitative behavior.
\section{Results}\label{sec:results}
In this section we present the main results for the shot noise and the current through the two level system that is modeled by Eq. (\ref{eq:model_Hamiltonian}). The calculations are based on the rate equations method which we presented in Sec. \ref{sec:rate equation formalism}. The results are explained using the approximated intuitive approach of signal analysis that we developed in Sec. \ref{sec:qualitative description}.

\subsection{Strong interactions}
\begin{figure}
\includegraphics[bb=10bp 109bp 585bp 691bp,clip,scale=0.425]{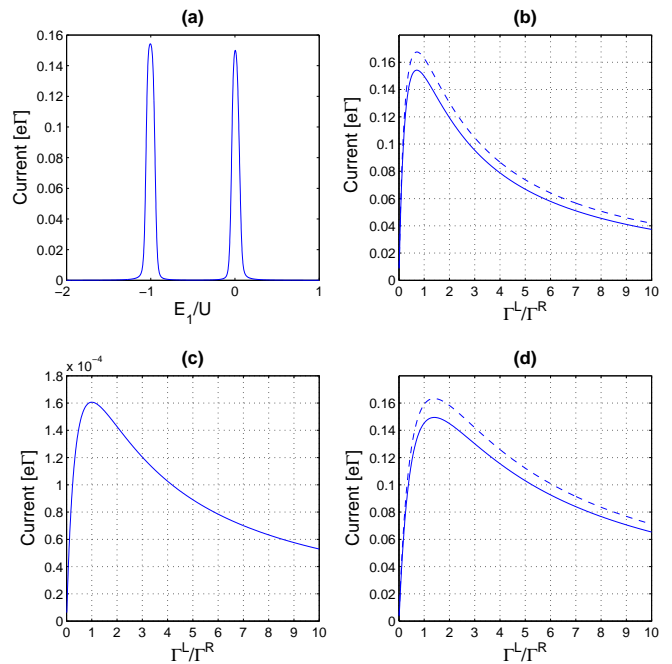}
\caption{The current through the two level system with the following parameters ($\Gamma\equiv\Gamma^R_1+\Gamma^L_1+\Gamma^R_2+\Gamma^L_2$): $K_BT=\hbar\Gamma$, $eV_{\rm bias}=10\hbar\Gamma$, $E_1-E_2=\hbar\Gamma$, $U=100\hbar\Gamma$ and $\Gamma_1^{R,L}=10\Gamma_2^{R,L}$. We assume the same left-right asymmetry for the two levels: $\Gamma^L_1/\Gamma^R_1=\Gamma^L_2/\Gamma^R_2\equiv\Gamma^L/\Gamma^R$. \textbf{(a)} The symmetric ($\Gamma^L/\Gamma^R=1$) current as a function of $E_1$. \textbf{(b)} The current at the right peak, $E_1=0$, as a function of the left-right asymmetry. The dashed line is the sequential tunneling current, and the solid line is the current including both sequential and cotunneling processes. \textbf{(c)} The (cotunneling) current at the valley, $E_1=-U/2$, as a function of the left-right asymmetry. The sequential tunneling current is practically zero at the valley. \textbf{(d)} The current at the left peak, $E_1=-U$, as a function of the left-right asymmetry. The dashed line is the sequential tunneling current, and the solid line is the current including both sequential and cotunneling processes.}
\label{fig: currents}
\end{figure}
{\it The current}: In Fig. \ref{fig: currents}(a), we plot the symmetric current \mbox{($\Gamma_i^L=\Gamma_i^R$)} through the two level system as a function of $E_1$ at large $U$ ($U>eV_{\rm bias}$) with the following parameters ($\Gamma\equiv\Gamma^R_1+\Gamma^L_1+\Gamma^R_2+\Gamma^L_2$): $K_BT=\hbar\Gamma$, $eV_{\rm bias}=10\hbar\Gamma$, \mbox{$E_1-E_2=\hbar\Gamma$}, $U=100\hbar\Gamma$ and $\Gamma_1^{R,L}=10\Gamma_2^{R,L}$. Two Coulomb peaks with a width $\sim eV_{\rm bias}$ appear in the current: at $E_1\approx0$ (the right peak), where the two levels are between the leads' chemical potentials, $\mu^L$ and $\mu^R$ (similar to the schematic picture in Fig. \ref{fig:system}), and at $E_1\approx-U$, where one of the levels is occupied so the other level is effectively shifted and placed between $\mu^L$ and $\mu^R$. The right peak is a bit lower than the left peak because of the small energy difference between the two levels ($E_1>E_2$). At finite temperature, this small energy difference makes the probability of finding level 2 occupied a bit larger than the probability of finding level 1 occupied. While near the right peak the current through the strongly coupled level- level 1- is blocked by the occupation of level 2, near the left peak the occupation of level 2 allows it, and therefore the left peak is a bit higher. If the sign of the energy difference was the opposite ($E_1<E_2$) the right peak was higher than the left peak.

In Figs. \ref{fig: currents}b and \ref{fig: currents}d we plot the left-right asymmetry dependence of the current at the right and left peaks respectively. We assume the same left-right asymmetry for the two levels, $\Gamma^L_1/\Gamma^R_1=\Gamma^L_2/\Gamma^R_2\equiv\Gamma^L/\Gamma^R$. Due to the lack of a particle-hole symmetry at the peaks the current is not maximal when the system is symmetrically coupled to the leads. To understand this consider the current at the left peak in the simple case of zero temperature and no cotunneling processes. The right lead current (\ref{eq0}) in this case is simply $\langle I\rangle=P_{11}(\Gamma_1^R+\Gamma_2^R)\approx P_{11}\Gamma_1^R$ where $P_{11}$ is the probability of finding the system doubly occupied. The probability $P_{11}$ can be easily calculated since the probability of finding the system empty in this case (zero temperature and $E_1=-U$) is zero. Also, in the limit $|E_1-E_2|\ll V_{\rm bias}$ the probability of finding only level 1 occupied, $P_{10}$, and the probability of finding only level 2 occupied $P_{01}$ are identical and given by $P_{\rm e}$ of Eq. (\ref{eq: solution of Pe}). The probability of finding the system doubly occupied is therefore $P_{11}=1-2P_{\rm e}$. The average current is $\langle I\rangle=\Gamma_1^R(1-2P_{\rm e})=(\Gamma^L_1+\Gamma^R_1)\frac{1}{\Gamma^L/\Gamma^R+1}(1-\frac{2}{\Gamma^L/\Gamma^R+2})$, and it is maximal, for a fixed $\Gamma^L_1+\Gamma^R_1$, at $\Gamma^L/\Gamma^R=\sqrt{2}$. Indeed, the current in Fig. \ref{fig: currents}d is maximal at $\Gamma^L/\Gamma^R\approx\sqrt{2}$ (the calculation is done at finite temperature and includes cotunneling therefore $\sqrt{2}$ is only an approximation). Similarly, the maximum of the right peak is at $\Gamma^L/\Gamma^R\approx1/\sqrt{2}$. The current at the valley $E_1=-U/2$, depicted in Fig. \ref{fig: currents}c, carried by cotunneling processes, is maximal where the two levels are symmetrically coupled to the leads, $\Gamma^L=\Gamma^R$.

{\it The Fano factor}. The current is carried by tunneling of electrons, namely current pulses of charge 'e', resulting in a shot noise. On top of the tunneling events, as we discussed in Sec. \ref{sec:qualitative description}, there is also a telegraph noise; by tunneling into or out of level 2, we change the tunneling rate through level 1. Since most of the current is carried by tunneling through level 1 the current alternates between two different average values. For example, if we focus on the left peak of the current, $E_1=-U$ (see Fig. \ref{fig: currents}a), most of the tunneling events are via level 1 while level 2 is occupied. However, because of the strong interactions, each time the electron leaves the narrow level, level 2, the current drops dramatically and resumes only when a new electron enters level 2.
\begin{figure}
\includegraphics[bb=113bp 118bp 508bp 683bp,clip,scale=0.61]{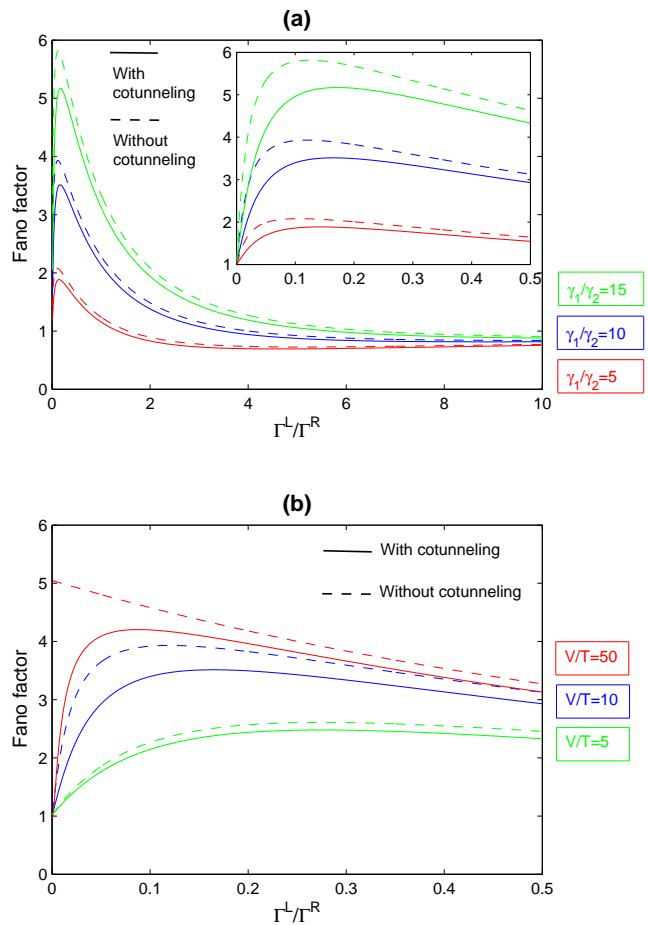}
\caption{The Fano factor, $F=\frac{S}{2e\langle I\rangle}$ at $E_1=-U$ (the left peak of the current) as a function of the left-right asymmetry with the following parameters: $eV_{\rm bias}=0.1U$, $E_1-E_2=0.01U$ and $\hbar\Gamma^L_1+\hbar\Gamma^R_1+\hbar\Gamma^L_2+\hbar\Gamma^R_2=0.01U$. We assume that the two levels have the same left-right asymmetry, $\Gamma^L/\Gamma^R$. When we change $\Gamma^L/\Gamma^R$ we keep the total width of each level fixed. \textbf{(a)} The Fano factor for different ratios of the two levels' widths ($\gamma_i=\hbar(\Gamma_i^R+\Gamma_i^L)/2$) with $K_BT=0.01U$ and $eV_{\rm bias}=0.1U$. The dashed lines are the sequential tunneling Fano factors and the solid lines are the calculated Fano factors including cotunneling processes. In the inside box in the upper right corner we zoom in on the small values of $\Gamma^L/\Gamma^R$. The Fano factor  is larger than one (super-Poissonian noise) and maximal in asymmetric coupling. \textbf{(b)} The Fano factor as a function of the left-right asymmetry for different values of $eV_{\rm bias}/T$. The ratio between the levels' widths is $\Gamma^{R,L}_1=10\Gamma^{R,L}_2$. The dashed lines are the sequential tunneling Fano factors and the solid lines are the calculated Fano factors including cotunneling processes. The cotunneling processes dramatically decrease the Fano factor for large $eV_{\rm bias}/T$.}
\label{fig: fano}
\end{figure}

n Fig. \ref{fig: fano}, we depict the Fano factor at the left peak ($E_1=-U$) of the current through the two level system with the following parameters: $eV_{\rm bias}=0.1U$, $E_1-E_2=0.01U$, and $\hbar\Gamma^L_1+\hbar\Gamma^R_1+\hbar\Gamma^L_2+\hbar\Gamma^R_2=0.01U$. By increasing $\gamma_1/\gamma_2$, we increase the average number of electrons that tunnel through level 1 while level 2 is occupied (the quantity $\Gamma_1\tau_1$ in Eq. (\ref{eq:Fano_simplified_model})). Indeed, similar to what we expect for the simplified model of Sec. \ref{sec:qualitative description} (Eqs. (\ref{eq:Fano_simplified_model}) and (\ref{eq:Fano_simplified_zero_T})), the noise is larger for large $\gamma_1/\gamma_2$ as one can see from Fig. \ref{fig: fano}(a).

In Fig. \ref{fig: fano}, we plot the left-right asymmetry dependence of the Fano factor. Similar to the simplified case of Sec. \ref{sec:qualitative description}, we find the following asymmetry dependence of the Fano factor: for large $\Gamma^L/\Gamma^R$, the Fano factor is $F=1$, and as we decrease $\Gamma^L/\Gamma^R$, we see enhancement of the the Fano factor. Also, as we discussed in Sec. \ref{sec:qualitative description}, the finite temperature suppresses the Fano factor below some $\Gamma^L/\Gamma^R\ll1$ toward the value $F=1$ in the limit $\Gamma^L/\Gamma^R\rightarrow0$. The most interesting feature in Fig. \ref{fig: fano} is the unexpected influence that the cotunneling processes have on the Fano factor. At low temperatures ($K_BT\ll eV_{\rm bias}$), the cotunneling processes suppress the Fano factor significantly in the asymmetric coupling regime $\Gamma^L\ll\Gamma^R$ (the dashed lines in Fig. \ref{fig: fano} represent calculations without cotunneling processes).

The physics behind the reduction of the Fano factor in the $\Gamma^L/\Gamma^R\ll1$ limit can be explained by the simplified picture of Sec. \ref{sec:qualitative description}. At zero temperature and taking into account sequential tunneling only, as we reduce $\Gamma^L/\Gamma^R$ the Fano factor is enhanced toward the value $F\approx\frac{\gamma_1}{2\gamma_2}$ in the limit $\Gamma^L/\Gamma^R\rightarrow0$ (see Eqs. (\ref{eq:Fano_simplified_zero_T}) and (\ref{eq: solution of Pe})). This suits the upper (red) dashed line in Fig. \ref{fig: fano}. Finite temperature allows backward tunneling- opposite to the voltage bias direction, which reduces the number of current pulses through level 1 each time level 2 is occupied. The reason is that for $\Gamma^L\ll\Gamma^R$ electrons tunnel many times back and forth between the right lead and the system before a tunneling event from the left lead to the system takes place. Thus, the quantity $\Gamma_1\tau_1$ in Eq. (\ref{eq:Fano_simplified_model}) (see also Fig. \ref{fig:qualitative}) is reduced and the Fano factor is suppressed.

Cotunneling processes, similar to the finite temperature, allow backward tunneling. Virtual tunneling to the left is possible as a part of a full two particle cotunneling process. The important cotunneling processes which suppress the Fano factor in the asymmetric limit are the inelastic processes than change the system's state between the states $|1,0\rangle\longleftrightarrow|0,1\rangle$. In the asymmetric limit, the total rate of these processes becomes larger than the rate of sequential tunneling from the left lead into level 1. As a result, the occupation of level 2 changes faster than the time delay between consecutive current pulses through level 1. The quantity $\Gamma_1\tau_1$ is reduced and the Fano factor is suppressed toward the value $F=1$ in the extremely asymmetric limit $\Gamma^L/\Gamma^R\rightarrow0$.

We want to mention the experimental work that was reported in Ref. [\onlinecite{PhysRevLett.98.066801}], where a super-Poisson noise with a strong asymmetry dependence was measured in quantum dot that was attached to two leads at strong magnetic field. One of the suggested explanations for the enhancement of the noise in this system was an additional level that is weakly coupled to the leads. Indeed, the strong dependence of the Fano factor on the asymmetry of the dot-leads coupling, which is very similar to the asymmetry dependence of the Fano factor we have in Fig. \ref{fig: fano}, indicates a possible two level system. We must emphasize that although our results fits qualitatively the experimental measurements \cite{footnote_on_experiment}, the rate equations formalism is not suitable for quantitative analysis of the experimental results since the quantum dot was attached relatively strong to the leads, a situation that makes the rate equations approach invalid. Nevertheless, qualitatively, we believe that the significant asymmetry dependence of the Fano factor is a strong evidence for the presence of a second interacting level that was weakly attached to the leads in the experimental setup.

\emph{Spinful electrons}. In order to avoid unnecessary complexities, we have assumed that the electrons are spinless. Physically, this situation can be realized by two-level quantum dot at strong magnetic field or two single level quantum dots at strong magnetic field. The spin degree of freedom can be added to the problem in two ways: single spinful level with spin dependent couplings to the leads or two spinful levels with different couplings to the leads. The single spinful level case is very similar to the system that we analyze, the only difference is the fact that processes with different spins cannot interfere. Therefore some of the cotunneling rates are slightly different (\emph{e.g.}, cotunneling of electron through an empty level). Nevertheless, we want to stress that the most important cotunneling processes in the more physical situation of relatively strong interactions, which are the inelastic processes $|1,0\rangle\leftrightarrow|0,1\rangle$ (or $|\uparrow\rangle\leftrightarrow|\downarrow\rangle$), have exactly the same rates as in our model. Moreover, and this is the important point, the physics behind the noise enhancement in the spinful case and in the spinless case, is the same. The noise is enhanced because of the blocking effect which is a result of the Coulomb interactions. Therefore, the same qualitative dependence on the asymmetry is expected.
This is true also in the two spinful levels case, and in fact also in multi-level systems with interactions. If the blocking effect enhances the noise, we expect similar dependence of the Fano factor on $\Gamma^L/\Gamma^R$ with similar suppression due to finite temperature and cotunneling.

\subsection{Weak interactions}
\begin{figure}
\includegraphics[bb=10bp 100bp 585bp 697bp,clip,scale=0.42]{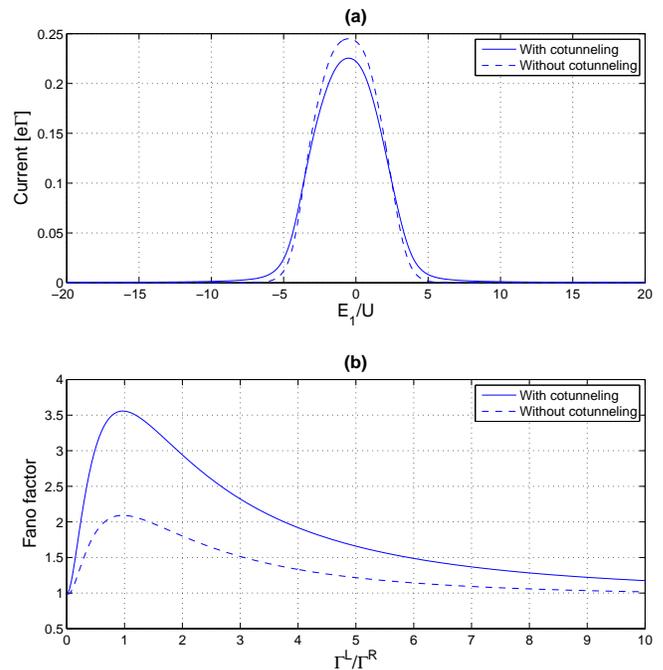}
\caption{The current and the Fano factor in the small U ($U<eV_{\rm bias}$) region with the following parameters ($\Gamma\equiv\Gamma^R_1+\Gamma^L_1+\Gamma^R_2+\Gamma^L_2$): $U=2\hbar\Gamma$, $eV_{\rm bias}=10\hbar\Gamma$, $K_BT=\hbar\Gamma$, $E_1-E_2=0.5\hbar\Gamma$ and $\gamma_1=10000\gamma_2$. \textbf{(a)} The current as a function of $E_1$ at the symmetric point, $\Gamma^L=\Gamma^R$. The dashed line is the current including sequential tunneling only, and the solid line is the current including also cotunneling processes. \textbf{(b)} The Fano factor at $E_1=0$ as a function of the left-right asymmetry. The dashed line is the sequential tunneling Fano factor and the solid line is the Fano factor including cotunneling processes. The cotunneling processes enhance the Fano factor significantly.}
\label{fig: smallU}
\end{figure}

While $U$ is large, say $U>eV_{\rm bias}$, the current at the left (right) peak, $E_1\approx-U$ ($E_1\approx0$) is changed dramatically when an electron is tunneling into or out of level 2, since the strong interaction blocks the tunneling through level 1.  This is not the case for small $U$. If $U$ is small compared to $V_{\rm bias}$ the current alternates between two relatively close values, and the effect on the noise is smaller. Yet, if the ratio between the levels' widths, $\gamma_1/\gamma_2$ is large enough, the enhancement of the noise can be important. In Fig. \ref{fig: smallU} we depict the current and the Fano factor for relatively small $U$ ($U=eV_{\rm bias}/5$) with the following parameters ($\Gamma\equiv\Gamma^R_1+\Gamma^L_1+\Gamma^R_2+\Gamma^L_2$): $U=2\hbar\Gamma$, $eV_{\rm bias}=10\hbar\Gamma$, $K_BT=\hbar\Gamma$, $E_1-E_2=0.5\hbar\Gamma$ and $\gamma_1=10000\gamma_2$. We see again that the cotunneling processes are important as they enhance the Fano factor significantly (the cotunneling processes contribute less than 20\% of the current but almost double the Fano factor). Notice also that unlike the Fano factor of the strong $U$ case, the Fano factor is maximal when the system is symmetrically coupled to the leads, $\Gamma^L=\Gamma^R$.
\begin{figure}
\includegraphics[bb=15bp 113bp 586bp 687bp,clip,scale=0.43]{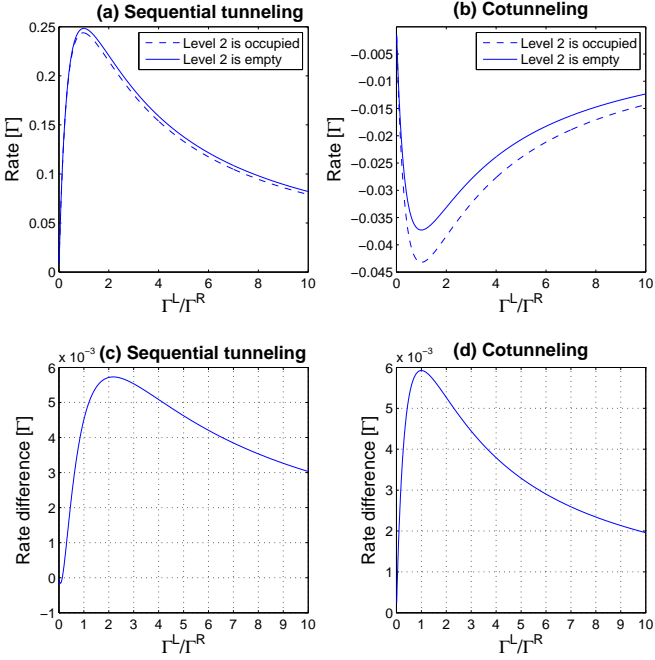}
\caption{The dependence of the tunneling rates through level 1 on the occupation of level 2 in the small U ($U<eV_{\rm bias}$) region with the following parameters ($\Gamma\equiv\Gamma^R_1+\Gamma^L_1+\Gamma^R_2+\Gamma^L_2$): $U=2\hbar\Gamma$, $V_{\rm bias}=10\hbar\Gamma$, $K_BT=\hbar\Gamma$, $E_1=0$, $E_2=-0.5\hbar\Gamma$ and $\gamma_1=10000\gamma_2$. Negative cotunneling rates mean that the cotunneling processes decrease the current (see for example Ref. [\onlinecite{PhysRevB.74.205438}]). \textbf{(a)} The sequential tunneling rate from the left lead through level 1 to the right lead ($((\omega_{11,01}^{\rightarrow})^{-1}+(\omega_{01,11}^{\rightarrow})^{-1})^{-1}$ when level 2 is occupied and $((\omega_{10,00}^{\rightarrow})^{-1}+(\omega_{00,10}^{\rightarrow})^{-1})^{-1}$ when level 2 is empty). \textbf{(b)} The cotunneling rates from the left lead through level 1 to the right lead ($\omega_{11,11}^{\rightarrow}+\omega_{01,01}^{\rightarrow}$ when level 2 is occupied and $\omega_{10,10}^{\rightarrow}+\omega_{00,00}^{\rightarrow}$ when level 2 is empty). \textbf{(c)} The difference between the sequential tunneling rates of the two stages (level 2 is empty and full). \textbf{(d)} The difference between the cotunneling rates of the two stages (level 2 is empty and full).}
\label{fig:rates_weak_U}
\end{figure}

To explain the enhancement of the noise in the small $U$ case, we use the intuitive picture that we studied in Sec. \ref{sec:qualitative description}; a sequence of current pulses through level 1 and a random telegraph signal describing the occupation of level 2. The rate of the pulses depends on the occupation of level 2 and we consider two different rates: $\Gamma_1$ describes the tunneling rate through level 1 while level 2 is full and $\tilde{\Gamma}_1$ describes the tunneling rate while level 2 is empty. The Fano factor is given by (\ref{eq:Fano_simplified_smallU_approx})
\begin{equation}
F\approx 1+\frac{2(\Delta\Gamma_1)^2}{\Gamma_1\Gamma_2}P_{\rm e}^2(1-P_{\rm e})^2,
\end{equation}
where the rate $\Gamma_2=(\tau_0+\tau_1)^{-1}$ is the tunneling rate through level 2, $\Delta\Gamma_1\equiv\tilde{\Gamma}_1-\Gamma_1$, $P_{\rm e}=\frac{\tau_0}{\tau_0+\tau_1}$, and we have used the fact that $\tilde{\Gamma}_1=\Gamma_1+\Delta\Gamma_1\approx\Gamma_1$ for $U<eV_{\rm bias}$. We see that although the Fano factor is usually reduced to one for $\frac{\Delta\Gamma_1}{\Gamma_1}\ll1$, in the extreme case of $\frac{\Gamma_2}{\Gamma_1}<\left(\frac{\Delta\Gamma_1}{\Gamma_1}\right)^2$ the Fano factor can be enhanced.

We can now understand why the Fano factor is enhanced noticeably due to the cotunneling processes. For $U<eV_{\rm bias}$, the rate of the sequential tunneling through level 1 barely depends on the occupation of level 2 since the energies $E_1$ and $E_1+U$ are close to each other compared to $V_{\rm bias}$. The cotunneling processes however, are much more sensitive to $U$. In Fig. \ref{fig:rates_weak_U} we illustrate this point by comparing the dependence of the sequential tunneling and the cotunneling rates through level 1 on the occupation of level 2, with the same parameters of Fig. \ref{fig: smallU}b. In Fig. \ref{fig:rates_weak_U}(a), we plot the sequential tunneling rates from the left lead through level 1 to the right lead when level 2 is empty- $((\omega_{10,00}^{\rightarrow})^{-1}+(\omega_{00,10}^{\rightarrow})^{-1})^{-1}$, and full- $((\omega_{11,01}^{\rightarrow})^{-1}+(\omega_{01,11}^{\rightarrow})^{-1})^{-1}$ (expressions for the rates are given in Appendix \ref{app: list of rates}, see Fig. \ref{fig: sequential processes} to clarify the notations). In Fig. \ref{fig:rates_weak_U}b we plot the cotunneling rates from the left lead through level 1 when level 2 is empty- $\omega_{10,10}^{\rightarrow}+\omega_{00,00}^{\rightarrow}$, and full- $\omega_{11,11}^{\rightarrow}+\omega_{01,01}^{\rightarrow}$ (expressions for the rates are given in Appendix \ref{app: list of rates}, see Fig. \ref{fig: elastic_cotunneling_processes} to clarify the notations). In Figs. \ref{fig:rates_weak_U}c and \ref{fig:rates_weak_U}d we plot the difference between the values of the tunneling rates through level 1 when level 2 is empty and the values of the tunneling rates when level 2 is full (in Fig. \ref{fig:rates_weak_U}c the sequential tunneling rate and \ref{fig:rates_weak_U}d the cotunneling rate). We see that although the sequential tunneling rate is order of magnitude larger than the cotunneling rate, the cotunneling processes are much more sensitive to the occupation of level 2, making the difference between the two values of the cotunneling rate on the same order of the difference between the two values of the sequential tunneling rate. Thus, even though the cotunneling processes have a small contribution to the current, they contribute the same as the sequential tunneling to $\Delta\Gamma_1$ and therefore have an important contribution to the Fano factor. We want to emphasize that the Fano factor is enhanced in the weak interactions regime only if level 2 is coupled to the leads extremely weaker than level 1, as we require $\frac{\Gamma_2}{\Gamma_1}<\left(\frac{\Delta\Gamma_1}{\Gamma_1}\right)^2$ where $\left(\frac{\Delta\Gamma_1}{\Gamma_1}\right)^2$ is a very small number.

Since (\ref{eq:Fano_simplified_smallU_approx}) depends quadratically on the multiplication of the probabilities of finding level 2 empty and full, $P_{\rm e}(1-P_{\rm e})$, the enhancement of the Fano factor due to $\Delta\Gamma_1$ is maximal where $P_{\rm e}(1-P_{\rm e})$ is maximal. In the limit of small $U$, the energy of the system, wether it is empty, singly occupied or doubly occupies, is more or less the same for $E_1=0$, therefore $P_{\rm e}(1-P_{\rm e})$ is maximal where the system is symmetrically coupled to the leads.
\section{Conclusions}\label{sec: conclusions}
We have analyzed the transport through a two level system that is coupled to two leads with one level coupled much stronger than the other. We showed that a simple intuitive model that describes the transport through this system as a sequence of current pulses that is stopped occasionally captures well most of the transport properties of the system. The current pulses in this model correspond to tunneling events through the strongly coupled level, that are stopped occasionally by tunneling events through the weakly coupled level. As expected from this simple model, we find in the more rigorous rate equations based calculation, a super-Poisson noise, with Fano factor larger than one, indicating that the electrons tunnel in bunches.

We showed, using rate equations calculation, a unique dependence of the transport on the asymmetry of the coupling to the leads, $\Gamma^L/\Gamma^R$, in the strong interaction limit (see Figs \ref{fig: currents} and \ref{fig: fano}). In particular, the Fano factor is enhanced at asymmetric couplings and increases as we decrease $\Gamma^L/\Gamma^R$. Eventually, the finite temperature suppresses the Fano factor in the limit $\Gamma^L/\Gamma^R\rightarrow0$.

Interestingly, at $eV_{\rm bias}\gg K_BT$ the cotunneling processes affect dramatically the enhancement of the Fano factor. The cotunneling processes that change the system's state between the states $|1,0\rangle$ and $|0,1\rangle$ (level 1 occupied and level 2 occupied) suppress the Fano factor even for $K_BT/eV_{\rm bias}\ll1$ and therefore they are essential for transport calculations even though most of the current is carried by sequential tunneling processes. Other cotunneling processes are less critical in the strong interaction regime, they slightly improve the approximation but don't change the transport significantly.

At relatively weak interactions, we showed that the Fano factor may also be enhanced noticeably if the difference between the couplings of the two levels is extremely large, namely $\gamma_2/\gamma_1\ll1$. In this limit we find again that the cotunneling processes are essential for the analysis of the noise and the Fano factor. The reason is because the telegraphic fluctuations of the sequential processes are of the same order as the fluctuations of the cotunneling processes. Thus, even though the current is described by sequential tunneling quite well, one must include the cotunneling processes in the noise calculations.

\section*{ACKNOWLEDGMENTS}
We would like to thank A. Amir and E. Shahmoon for useful discussions. This work was supported by the BSF and the DIP.
\appendix
\section{List of tunneling rates}\label{app: list of rates}
In this appendix we give the expressions for all the tunneling rates that enter the rate equations.
\subsection*{Sequential tunneling rates}
To the lowest order in $H_{{\rm t}}$ the transition rates can be calculated using Fermi's golden rule. We use the notation $\omega_{i,j}^{\rightarrow}$ for the rate of a tunneling process that changes the system's state from '$i$' to '$j$' by tunneling an electron from the left to the right direction (and similarly $\omega_{i,j}^{\leftarrow}$ for electron that moves from right to left). For example $\omega_{00,10}^{\rightarrow}$ is the rate of tunneling from the left lead into level 1 while level 2 is empty. The rates of the sequential tunneling processes are
\begin{eqnarray}
&&\omega_{00,10}^{\rightarrow}=\Gamma^L_1 F_{FD}(E_1-\mu^L),\\
&&\omega_{00,01}^{\rightarrow}=\Gamma^L_2 F_{FD}(E_2-\mu^L),\\
&&\omega_{10,00}^{\rightarrow}=\Gamma^R_1 F_{FD}(\mu^R-E_1),\\
&&\omega_{01,00}^{\rightarrow}=\Gamma^R_2 F_{FD}(\mu^R-E_2),\\
&&\omega_{01,11}^{\rightarrow}=\Gamma^L_1 F_{FD}(E_1+U-\mu^L),\\
&&\omega_{10,11}^{\rightarrow}=\Gamma^L_2 F_{FD}(E_2+U-\mu^L),\\
&&\omega_{11,01}^{\rightarrow}=\Gamma^R_1 F_{FD}(\mu^R-E_1-U),\\
&&\omega_{11,10}^{\rightarrow}=\Gamma^R_2 F_{FD}(\mu^R-E_2-U).
\end{eqnarray}
We have defined
\begin{equation}
\Gamma^L_i\equiv\nu\frac{2\pi}{\hbar}|t_i^L|^2,\mbox{\ \ } \Gamma^R_i\equiv\nu\frac{2\pi}{\hbar}|t_i^R|^2,
\end{equation}
and the Fermi's function $F_{FD}(x)=(1+e^{\beta x})^{-1}$ gives the probability for the availability of an electron or a hole for the tunneling process. The left moving rates, $\omega_{i,j}^{\leftarrow}$, have similar expressions with $\mu^L\leftrightarrow\mu^R$ and $\Gamma_{1,2}^L\leftrightarrow\Gamma_{1,2}^R$.
\subsection*{Elastic-cotunneling rates}
Each elastic cotunneling process has two possible intermediate states. For example, electron can tunnel through an empty system via level 1 or 2, thus, cotunneling processes of the form $|0,0\rangle\longrightarrow|0,0\rangle$ have two possible intermediate states: $|1,0\rangle$ and $|0,1\rangle$. The two possible paths interfere and we need to sum the amplitudes of the two possibilities rather than their probabilities. We use the notation $\omega_{i,i}^{\rightarrow}$ ($\omega_{i,i}^{\leftarrow}$) for elastic-cotunneling processes in which the electron tunnels to the right (left) direction. The total elastic-cotunneling rates are the sum of the rates of all the possible processes, namely integrating over all incoming electron's energies.
\begin{widetext}
\begin{eqnarray}
\omega_{00,00}^{\rightarrow}&=&\frac{2\pi\nu^2}{\hbar}\int d\epsilon F_{FD}\left(\epsilon-\mu^L\right)\left(1-F_{FD}\left(\epsilon-\mu^R\right)\right)\left|\frac{t^L_1t^R_1}{\epsilon-E_1}+\frac{t^L_2t^R_2}{\epsilon-E_2}\right|^2,\label{eq:elastic00}\\
\omega_{10,10}^{\rightarrow}&=&\frac{2\pi\nu^2}{\hbar}\int d\epsilon F_{FD}\left(\epsilon-\mu^L\right)\left(1-F_{FD}\left(\epsilon-\mu^R\right)\right)\left|\frac{t^L_1t^R_1}{\epsilon-E_1}+\frac{t^L_2t^R_2}{\epsilon-E_2-U}\right|^2,\label{eq:elastic10}\\
\omega_{01,01}^{\rightarrow}&=&\frac{2\pi\nu^2}{\hbar}\int d\epsilon F_{FD}\left(\epsilon-\mu^L\right)\left(1-F_{FD}\left(\epsilon-\mu^R\right)\right)\left|\frac{t^L_1t^R_1}{\epsilon-E_1-U}+\frac{t^L_2t^R_2}{\epsilon-E_2}\right|^2,\label{eq:elastic01}\\
\omega_{11,11}^{\rightarrow}&=&\frac{2\pi\nu^2}{\hbar}\int d\epsilon F_{FD}\left(\epsilon-\mu^L\right)\left(1-F_{FD}\left(\epsilon-\mu^R\right)\right)\left|\frac{t^L_1t^R_1}{\epsilon-E_1-U}+\frac{t^L_2t^R_2}{\epsilon-E_2-U}\right|^2\label{eq:elastic11}.
\end{eqnarray}
\end{widetext}
The left moving elastic-cotunneling rates, $\omega_{i,i}^{\leftarrow}$, have similar expressions with $\mu^L\leftrightarrow\mu^R$. Equations (\ref{eq:elastic00})-(\ref{eq:elastic11}) are formal expressions, and the actual rates, which we use as input for the rate equations, cannot be directly calculated from these integrals. The reason is the divergence of these expressions due to the finite widths of the two levels (which we treat as two delta functions in energy). We use a regularization scheme \cite{Averin1994979,PhysRevB.65.115332} for the calculation of the cotunneling rates. The regularization scheme is summarized in Appendix \ref{app:regularization}.

\subsection*{Inelastic-cotunneling rates}
We consider the inelastic-cotunneling processes that change the system's state between $|1,0\rangle\longleftrightarrow|0,1\rangle$ and $|0,0\rangle\longleftrightarrow|1,1\rangle$. We begin with the former; we use the notation $\omega_{i,j}^{\rightleftarrows}$ for the rate of processes in which the system changes its state from $|i\rangle$ to $|j\rangle$ in the following way: the electron that enters or leaves level 1 tunnels to the right direction, while the electron that enters or leaves level 2 tunnels to the left, and similarly, we define $\omega_{i,j}^{\leftrightarrows},\omega_{i,j}^{\rightrightarrows},\omega_{i,j}^{\leftleftarrows}$. For example, if the system's initial state is $|1,0\rangle$ and the electron in level 1 tunnels to the right lead, while another electron from the right lead tunnels to level 2, we denote the rate of this process by $\omega_{10,01}^{\rightleftarrows}$. Again, there are two possible intermediate states for the processes $|1,0\rangle\longleftrightarrow|0,1\rangle$, and we need to sum them properly. The formal expression for these rates are
\begin{widetext}
\begin{eqnarray}
\omega_{10,01}^{\rightrightarrows}&=&\frac{2\pi\nu^2}{\hbar}\int d\epsilon F_{FD}\left(\epsilon-\mu^L\right)\left(1-F_{FD}\left(\epsilon+(E_1-E_2)-\mu^R\right)\right)\left|\frac{t^R_1t^L_2}{\epsilon-E_2}-\frac{t^R_1t^L_2}{\epsilon-E_2-U}\right|^2,\label{eq:inelastic1}\\
\omega_{10,01}^{\leftleftarrows}&=&\frac{2\pi\nu^2}{\hbar}\int d\epsilon F_{FD}\left(\epsilon-\mu^R\right)\left(1-F_{FD}\left(\epsilon+(E_1-E_2)-\mu^L\right)\right)\left|\frac{t^L_1t^R_2}{\epsilon-E_2}-\frac{t^L_1t^R_2}{\epsilon-E_2-U}\right|^2,\label{eq:inelastic2}\\
\omega_{10,01}^{\rightleftarrows}&=&\frac{2\pi\nu^2}{\hbar}\int d\epsilon F_{FD}\left(\epsilon-\mu^R\right)\left(1-F_{FD}\left(\epsilon+(E_1-E_2)-\mu^R\right)\right)\left|\frac{t^R_1t^R_2}{\epsilon-E_2}-\frac{t^R_1t^R_2}{\epsilon-E_2-U}\right|^2,\label{eq:inelastic3}\\
\omega_{10,01}^{\leftrightarrows}&=&\frac{2\pi\nu^2}{\hbar}\int d\epsilon F_{FD}\left(\epsilon-\mu^L\right)\left(1-F_{FD}\left(\epsilon+(E_1-E_2)-\mu^L\right)\right)\left|\frac{t^L_1t^L_2}{\epsilon-E_2}-\frac{t^L_1t^L_2}{\epsilon-E_2-U}\right|^2\label{eq:inelastic4}.
\end{eqnarray}
\end{widetext}
To get these integrals we used the energy conservation: if the incoming electron (that enters level 2) has the energy $\epsilon$, the outgoing electron (that leaves level 1) must have the energy $\epsilon+E_1-E_2$. To get the rates: $\omega_{01,10}^{\rightrightarrows}$, $\omega_{01,10}^{\leftleftarrows}$, $\omega_{01,10}^{\rightleftarrows}$ and $\omega_{01,10}^{\leftrightarrows}$, we may write integrals like (\ref{eq:inelastic1})-(\ref{eq:inelastic4}) and exchange $t^{L,R}_1\leftrightarrow t^{L,R}_2$ and $E_1\leftrightarrow E_2$. Once again, the formal expressions (\ref{eq:inelastic1})-(\ref{eq:inelastic4}) need to be regularized in order to extract the input terms for the rate equations calculation (see Appendix \ref{app:regularization}).

The last rates that we discuss are the inelastic $|0,0\rangle\longleftrightarrow|1,1\rangle$ processes' rates. These processes are somewhat more complex than the other cotunneling processes as they have four possible intermediate states. We use the notation $\omega_{00,11}^{\rightrightarrows}$ ($\omega_{00,11}^{\leftleftarrows}$) and $\omega_{11,00}^{\rightrightarrows}$ ($\omega_{11,00}^{\leftleftarrows}$) for processes in which two electrons enter or leave the two levels by tunneling to the right (left). We use the notation $\omega_{00,11}^{\leftrightarrow}$, $\omega_{11,00}^{\leftrightarrow}$ for processes in which the two electrons enter or leave the two levels by tunneling one to the right and the other to the left. The rates for the $|0,0\rangle\longleftrightarrow|1,1\rangle$ inelastic cotunneling processes are
\begin{widetext}
\begin{eqnarray}
\omega_{00,11}^{\rightrightarrows}&=\frac{2\pi\nu^2}{\hbar}\int d\epsilon\frac{1}{2} F_{FD}\left(\epsilon-\mu^L\right)F_{FD}\left((-\epsilon+E_1+E_2+U)-\mu^L\right)\left|\frac{t^L_1t^L_2}{\epsilon-E_1}-\frac{t^L_1t^L_2}{\epsilon-E_1-U}-\frac{t^L_1t^L_2}{\epsilon-E_2}+\frac{t^L_1t^L_2}{\epsilon-E_2-U}\right|^2,&\label{eq:inelastic5}\\
\omega_{00,11}^{\leftleftarrows}&=\frac{2\pi\nu^2}{\hbar}\int d\epsilon\frac{1}{2} F_{FD}\left(\epsilon-\mu^R\right)F_{FD}\left((-\epsilon+E_1+E_2+U)-\mu^R\right)\left|\frac{t^R_1t^R_2}{\epsilon-E_1}-\frac{t^R_1t^R_2}{\epsilon-E_1-U}-\frac{t^R_1t^R_2}{\epsilon-E_2}+\frac{t^R_1t^R_2}{\epsilon-E_2-U}\right|^2,&\label{eq:inelastic6}\\
\omega_{00,11}^{\leftrightarrow}&=\frac{2\pi\nu^2}{\hbar}\int d\epsilon F_{FD}\left(\epsilon-\mu^L\right)F_{FD}\left((-\epsilon+E_1+E_2+U)-\mu^R\right)\left|\frac{t^L_1t^R_2}{\epsilon-E_1}-\frac{t^L_1t^R_2}{\epsilon-E_1-U}-\frac{t^R_1t^L_2}{\epsilon-E_2}+\frac{t^R_1t^L_2}{\epsilon-E_2-U}\right|^2\label{eq:inelastic7},&\\
\omega_{11,00}^{\rightrightarrows}&=\frac{2\pi\nu^2}{\hbar}\int d\epsilon\frac{1}{2} F_{FD}\left(\mu^L-\epsilon\right)F_{FD}\left(\mu^L-(-\epsilon+E_1+E_2+U)\right)\left|\frac{t^L_1t^L_2}{\epsilon-E_1}-\frac{t^L_1t^L_2}{\epsilon-E_1-U}-\frac{t^L_1t^L_2}{\epsilon-E_2}+\frac{t^L_1t^L_2}{\epsilon-E_2-U}\right|^2,&\label{eq:inelastic8}\\
\omega_{11,00}^{\leftleftarrows}&=\frac{2\pi\nu^2}{\hbar}\int d\epsilon\frac{1}{2} F_{FD}\left(\mu^R-\epsilon\right)F_{FD}\left(\mu^R-(-\epsilon+E_1+E_2+U)\right)\left|\frac{t^R_1t^R_2}{\epsilon-E_1}-\frac{t^R_1t^R_2}{\epsilon-E_1-U}-\frac{t^R_1t^R_2}{\epsilon-E_2}+\frac{t^R_1t^R_2}{\epsilon-E_2-U}\right|^2,&\label{eq:inelastic9}\\
\omega_{11,00}^{\leftrightarrow}&=\frac{2\pi\nu^2}{\hbar}\int d\epsilon F_{FD}\left(\mu^L-\epsilon\right)F_{FD}\left(\mu^R-(-\epsilon+E_1+E_2+U)\right)\left|\frac{t^L_1t^R_2}{\epsilon-E_1}-\frac{t^L_1t^R_2}{\epsilon-E_1-U}-\frac{t^R_1t^L_2}{\epsilon-E_2}+\frac{t^R_1t^L_2}{\epsilon-E_2-U}\right|^2\label{eq:inelastic10}.&
\end{eqnarray}
\end{widetext}
The factor 1/2 that appears in Eqs. (\ref{eq:inelastic5}), (\ref{eq:inelastic6}), (\ref{eq:inelastic8}) and (\ref{eq:inelastic9}) is due to the double counting of processes: by integrating over $\epsilon$ we sum both $\epsilon=\epsilon'$ and $\epsilon=-\epsilon'+E_1+E_2+U$, however these two processes are identical since in both cases when the two levels are empty the two electrons are in the same lead with energies $\epsilon'$ and $-\epsilon'+E_1+E_2+U$. Hence, as we double count each process we insert a factor of $1/2$. Notice also that for $\omega_{00,11}^{\rightrightarrows}$, $\omega_{00,11}^{\leftleftarrows}$, $\omega_{11,00}^{\rightrightarrows}$ and $\omega_{11,00}^{\leftleftarrows}$, by including the point $\epsilon=\frac{E_1+E_2+U}{2}$ in the integral, we include an impossible process, as the two electrons in the lead have the same energy ($\frac{E_1+E_2+U}{2}$). Nevertheless, this point contribute zero to the integrals, and therefore we have no problems with the formal expressions (\ref{eq:inelastic5}), (\ref{eq:inelastic6}), (\ref{eq:inelastic8}) and (\ref{eq:inelastic9}).
The integrals in (\ref{eq:inelastic5})-(\ref{eq:inelastic10}) need a regularization in order to extract the rates that we use in the rate equations, the regularization scheme appears in Appendix \ref{app:regularization}.
\section{APPENDIX: REGULARIZATION SCHEME}\label{app:regularization}
In this appendix we summarize the regularization procedure for the cotunneling rates \cite{Averin1994979,PhysRevB.65.115332}. All the cotunneling rates that appear in Appendix \ref{app: list of rates}, excluding the $|0,0\rangle\longleftrightarrow|1,1\rangle$ rates, can be brought to the form
\begin{small}
\begin{eqnarray}
&&I(A,B,E_a,E_b,\mu_1,\mu_2)=\frac{2\pi\nu^2}{\hbar}\times\label{eq:general_cotunneling_rate}\\
&&\int d\epsilon F_{FD}(\epsilon-\mu_1)(1-F_{FD}(\epsilon-\mu_2))\left|\frac{A}{\epsilon-E_a}+\frac{B}{\epsilon-E_b}\right|^2,\nonumber
\end{eqnarray}
\end{small}where the amplitudes $A,B$ are multiplications of two tunneling coefficients, and one might need to use the relation $F_{FD}(-\epsilon)=1-F_{FD}(\epsilon)$ in order to bring the expression of a specific cotunneling rate to this form. The integral (\ref{eq:general_cotunneling_rate}) diverges due to the finite widths of the energy levels \cite{Averin1994979,PhysRevB.65.115332}. We first add by hand a width to the levels
\begin{small}
\begin{eqnarray}
I(A,B,E_a,E_b,\mu_1,\mu_2)=\frac{2\pi\nu^2}{\hbar}\times&&\label{eq:general_cotunneling_rate_finite_width}\\
\int d\epsilon F_{FD}(\epsilon-\mu_1)(1-F_{FD}(\epsilon-&\mu_2))\left|\frac{A}{\epsilon-E_a+i\gamma}+\frac{B}{\epsilon-E_b+i\gamma}\right|^2.&\nonumber
\end{eqnarray}
\end{small}
Next, we solve the integral (\ref{eq:general_cotunneling_rate_finite_width}) and write the solution as a power series in $\gamma$. We extract the cotunneling rate by subtracting the $1/\gamma$ term and taking the limit $\gamma\rightarrow0$. We should emphasize that in general each level has its own width and the sign $\pm i\gamma$ depends on the process; we associate different signs for incoming and outgoing electrons. If one adds the widths properly, the divergent term ($1/\gamma$) has a physical meaning, and one can read the sequential tunneling rates from it. Nevertheless, these details are not important for the regularization procedure, and the finite values of the cotunneling rates are independent of the details of the regulator $\gamma$. We can write (\ref{eq:general_cotunneling_rate_finite_width}) as
\begin{widetext}
\begin{equation*}
I(A,B,E_a,E_b,\mu_1,\mu_2)=\frac{2\pi\nu^2}{\hbar}\left(I1(A,E_a,\mu_1,\mu_2)+I1(B,E_b,\mu_1,\mu_2)+I2(A,B,E_a,E_b,\mu_1,\mu_2)\right),
\end{equation*}
where
\begin{eqnarray*}
&&I1(A,E_a,\mu_1,\mu_2)=\int d\epsilon F_{FD}(\epsilon-\mu_1)(1-F_{FD}(\epsilon-\mu_2))\left|\frac{A}{\epsilon-E_a+i\gamma}\right|^2,\\
&&I2(A,B,E_a,E_b,\mu_1,\mu_2)=\int d\epsilon F_{FD}(\epsilon-\mu_1)(1-F_{FD}(\epsilon-\mu_2))2Re\left\{\frac{A}{\epsilon-E_a+i\gamma}\frac{B}{\epsilon-E_b-i\gamma}\right\}.
\end{eqnarray*}
The solutions of $I1$ and $I2$ can be written using digamma functions with complex variables ($\psi(z)$). $I1$ contains a divergent part
\begin{equation}
I1(A,E_a,\mu_1,\mu_2)=\frac{|A|^2N_B(\mu_2-\mu_1)}{\gamma}
Im\left\{\psi\left(\frac{1}{2}+\frac{\beta\gamma}{2\pi}+\frac{i\beta}{2\pi}(\mu_2-E_a)\right)-\psi\left(\frac{1}{2}+\frac{\beta\gamma}{2\pi}+\frac{i\beta}{2\pi}(\mu_1-E_a)\right)\right\},
\end{equation}
where $\beta=1/(K_BT)$ and $N_B(\mu_2-\mu_1)=(e^{\beta(\mu_2-\mu_1)}-1)^{-1}$. After the subtraction of the $1/\gamma$ term and taking the limit $\gamma\rightarrow0$:
\begin{equation}
I1(A,E_a,\mu_1,\mu_2)=|A|^2N_B(\mu_2-\mu_1)\frac{\beta}{2\pi}Im\left\{\psi'\left(\frac{1}{2}+\frac{i\beta}{2\pi}(\mu_2-E_a)\right)-\psi'\left(\frac{1}{2}+\frac{i\beta}{2\pi}(\mu_1-E_a)\right)\right\}.
\end{equation}
There is no divergence in $I2$, therefore we simply solve it and take the limit $\gamma\rightarrow0$
\begin{eqnarray}
&&I2(A,B,E_a,E_b,\mu_1,\mu_2)=AB\frac{N_B(\mu_2-\mu_1)}{E_a-E_b}\times\\
&&Re\left\{\psi\left(\frac{1}{2}+\frac{i\beta}{2\pi}(E_a-\mu_2)\right)-\psi\left(\frac{1}{2}+\frac{i\beta}{2\pi}(E_b-\mu_2)\right)-\psi\left(\frac{1}{2}+\frac{i\beta}{2\pi}(E_a-\mu_1)\right)+\psi\left(\frac{1}{2}+\frac{i\beta}{2\pi}(E_b-\mu_1)\right)\right\}.\nonumber
\end{eqnarray}
\end{widetext}

We can use the solution of (\ref{eq:general_cotunneling_rate}) to solve the rates of the processes $|0,0\rangle\longleftrightarrow|1,1\rangle$. Notice that the formal expressions for these rates (Eqs. (\ref{eq:inelastic5})-(\ref{eq:inelastic10})) contain four terms inside the absolute value. We may use the trivial identity
\begin{eqnarray*}
&&|a+b+c+d|^2=|a+b|^2+|a+c|^2\\
&&-|a-d|^2-|b-c|^2+|b+d|^2+|c+d|^2,
\end{eqnarray*}
to write Eqs. (\ref{eq:inelastic5})-(\ref{eq:inelastic10}) as sums of six terms of the form of (\ref{eq:general_cotunneling_rate}).


\begin{thebibliography}{40}
\bibitem[{foo({\natexlab{a}})}]{footnote_on_spin}
\bibinfo{note}{Adding the spin degree of freedom does not change the
  qualitative behavior of the system and not important for the understanding of
  the underlying physics. We shortly discuss the spinful case in section 5.}
\bibitem[{foo({\natexlab{b}})}]{footnote_on_experiment}
\bibinfo{note}{Notice the unusual definition of the Fano factor that was used
  in Ref. [\onlinecite{PhysRevLett.98.066801}], denoted with $F^*$, where the
  Fano factor was divided by one minus the transmission giving inherently a
  larger value for the (modified) Fano factor at more symmetric
  configurations.}
\end{thebibliography}
\end{document}